\documentclass[%
 reprint,
 amsmath,amssymb,
 aps,
]{revtex4-2}

\usepackage[utf8]{inputenc}
\usepackage{dcolumn}
\usepackage{bm}
\usepackage{amsmath,mathtools}
\usepackage{amsthm}
\usepackage{amssymb}
\usepackage{graphicx}
\usepackage{hyperref}
\usepackage{subcaption}
\usepackage{float} 

\usepackage{physics}
\usepackage[usenames,dvipsnames]{color}

\def\be{\begin{equation}}
\def\ee{\end{equation}}
\def\l{\left}
\def\r{\right}

\graphicspath{{plots/}}

\begin{document}

\title{Interior Quasinormal Modes and Strong Cosmic Censorship}
\author{Filipe S. Miguel}
\email{fsdrm2@cam.ac.uk}
\affiliation{
 Department of Applied Mathematics and Theoretical Physics, University of Cambridge, \\
 Wilberforce Road, Cambridge CB3 0WA, United Kingdom\\
}%

\date{\today}

\begin{abstract}
Recent work has argued that the strong cosmic censorship (SCC) conjecture is violated by near-extremal Reissner–Nordström de Sitter (RNdS) black holes but respected by Kerr-de Sitter black holes. It has also been shown that the conjecture is violated by near-extremal BTZ black holes. The latter result relies on a coincidence between ``exterior'' and ``interior'' quasinormal frequencies. If this coincidence were to occur also for RNdS or Kerr-dS then it would significantly modify the conclusions of earlier work. In this paper, it is demonstrated that this coincidence does not occur for RNdS or Kerr-dS and so the conclusions of the earlier work remain valid.
\end{abstract}


\maketitle

\tableofcontents

\section{Introduction}

General relativity is a deterministic theory that describes our universe at large scales. Nevertheless, there are solutions of Einstein equations, that are not fully specified by the initial conditions. In fact, for some spacetimes, we find Cauchy horizons. These are regions beyond which there are infinitely many possible solutions to Einstein equations that are consistent with the initial data. This is a violation of determinism, and an undesirable feature of the theory. To attenuate this issue, in the 1970's, Penrose put forward the Strong Cosmic Censorship (SCC) conjecture \cite{PenroseSCC}. In broad terms SCC is the statement that spacetimes with Cauchy horizons, come from fine tuned initial conditions. \textit{Generic} spacetimes, should be deterministic.\\

Several solutions of Einstein equations are known to admit Cauchy horizons. Famous examples in asymptotically flat spacetimes $(\Lambda = 0)$ are Kerr and Reissner–Nordström black holes. Fortunately, there is compelling evidence \cite{SimpsonInstabRN, McNamaraInstBHInnerH, ChandrasekharCrossingRNCH, IsraelInternalStruct, DafermosIntRNBHs,SCC_Kerr_BH_proof,SCC_RN_BH_proof} that these horizons are not stable. In fact signals emitted by an observer outside the black hole (BH), suffer an infinite blueshift when approaching the Cauchy horizon, prompting the instability. On the other hand, for asymptotically de-Sitter spacetimes $(\Lambda > 0)$, there is a competing redshift effect due to the existence of a cosmological horizon, which may weaken the strength of the Cauchy horizon instability and make a violation of SCC more likely \cite{MossRNdS}.\\

Recently, Cardoso et al. \cite{CardosoSCC}, renewed the interest on SCC in asymptotically de-Sitter spacetimes. The authors showed that if we couple the Einstein field equations with a massless scalar field, then for a non-trivial set of RNdS black holes it is possible to extend the scalar field across the Cauchy horizon as a weak solution of the Einstein Klein-Gordon equations. More formally, they found that for a generic class of RNdS black holes, the massless scalar field will have finite energy at the Cauchy horizon. This work was based in recent mathematical results \cite{HintzlinearWavesCH,CostaStabRNdS} about the boundedness of linear fields near the Cauchy Horizon. In \cite{Kyriakos2}, the authors extended this result for a charged scalar field. In these papers, the authors use \cite{Nonlinear1,Nonlinear2,Noninear3}, to argue that the results should remain valid in the non-linear case. Further work from Dias et al. \cite{HarveyRoughSmooth, HarveyRNdSchargedField} showed that in the case of gravitoelectromagnetic perturbations, the violation is even more drastic. For a non-trivial region of parameter space gravitoelectromagnetic perturbations can be extended across the Cauchy horizon with arbitrary regularity. \\

RNdS BHs are not very relevant as astrophysical objects. It is not very likely that a near extremal amount of electrical charge falls into a black hole, to reproduce the SCC violation found in \cite{HarveyRoughSmooth}. It would be much more interesting, if we found SCC violations on the more physical Kerr-dS background. However, in \cite{HarveySccKerrdS}, the authors showed that this is not the case. In fact, scalar field and gravitational perturbations are sufficiently irregular at the Cauchy Horizon to preserve SCC. \\

In a recent paper, \cite{HarveyBTZ} Dias et al. studied the behaviour of several fields propagating around the 3-dimensional, asymptotically AdS, BTZ black hole \cite{BTZpaper}. They found that for BHs sufficiently close to extremality, these fields can be continued across the inner event horizon with arbitrary regularity.  This conclusion relies fundamentally on an unexpected coincidence found between quasinormal mode (QNM) frequencies of waves propagating on the exterior of the BH and waves propagating on the interior of the BH. This coincidence increases the regularity of fields on the Cauchy horizon, leading to a violation of SCC.\\

If this coincidence also occurred for Kerr-dS or RNdS black holes, we could find new violations of SCC in 4 dimensional gravity. In \cite{HarveyRoughSmooth} the authors found a large class of near extremal RNdS BHs that violate SCC, so this equality would simply increase the region of parameter space that violates SCC. Much more interesting is perhaps the Kerr-dS case. As discussed above, there is no evidence for SCC violations in this spacetime. However, if this coincidence was to occur, the conclusion of \cite{HarveySccKerrdS} would have to be revised. In fact, for a set of near extremal (NE) black holes, we could expect SCC to be violated. \\

In this paper we prove that this is not the case. We study the spectrum of interior and exterior QNMs for Kerr-dS and RNdS BHs. Using numerical and analytical arguments, we prove that the frequencies are distinct, maintaining the conclusions of \cite{HarveyRoughSmooth} and \cite{HarveySccKerrdS}.\\

It is also important to mention recent efforts in the direction of restoring the faith in SCC. At the classical level, we should mention \cite{DafermosRoughDataSCC}. In this paper, Dafermos and Shlapentokh-Rothman showed that even though some notions of SCC are not respected for a class of RNdS black holes, if we allow non-smooth initial data, then SCC is recovered.  More recently, a couple of papers by Hollands et al. , \cite{WaldInstabCH,Hollands2} restored the faith in SCC, when considering quantum fields on a RNdS / Kerr-dS BH.  The authors argue that the energy momentum tensor of quantum fields is sufficiently irregular at the Cauchy horizon, recovering SCC. The BTZ case is a bit more complicated. In \cite{HarveyBTZ}, the authors prove that the stress energy of quantum fields is regular at the Cauchy horizon. Hence, the backreaction onto the gravitational sector, should also remain regular. However, in a recent paper \cite{EmparanSCCBTZ} Emparan et al. , argued that second order backreaction diverges at the Cauchy Horizon, preserving SCC.\\

The paper is organized as follows. In section \ref{sec:background}, we review the RNdS and Kerr-dS black hole solutions. Then, we review the notion of exterior QNMs, define interior QNMs and finally outline the argument that relates QNMs with SCC. Then in section \ref{sec:SCCRNdS} we  study the families of interior and exterior QNMs that are relevant for SCC violation in the RNdS black hole. Using analytical approximations, we show that in the near extremal limit exterior and interior QNM frequencies coincide to leading order. Then, using high precision numerical methods, we find the values of these frequencies, and show that they are different. We conclude, that in the RNdS case, there is no additional violation of SCC. In section \ref{sec:Kerr-dS} we use a WKB argument to prove that exterior QNM frequencies are not frequencies of interior QNMs. We conclude reinforcing confidence of SCC in this spacetime.

\section{Background Material \label{sec:background}}

\subsection{The RNdS black hole solution\label{subsec:RNdS}}

Let's review the 4 dimensional RNdS black hole (see e.g. \cite{WaldBook}). This solution of Einstein's equations describes an electrically charged black hole in a de-Sitter background. The BH is completely specified by 3 parameters, $(M, Q, \Lambda)$, (mass, charge and cosmological constant). In static coordinates $(t, r, \theta, \phi)$, the line element of the metric can be written as:
\begin{equation}
	\dd s^2\,=\,-\,f\,\dd t^2\,+\,\frac{\dd r^2}{f}\,+\,r^2\,\dd \Omega_2^2,
\end{equation}
where $\dd \Omega^2$ is the line element of a unit radius $S^2$, and
\begin{equation}
	f(r) \,= \, 1\,-\,\frac{2 M}{r}\,+\,\frac{Q^2}{r^2} \,-\, \frac{\Lambda}{3} \,r^2. \label{eq:f0}\\
\end{equation}

\begin{figure}
\centering
\includegraphics[width=0.9\linewidth]{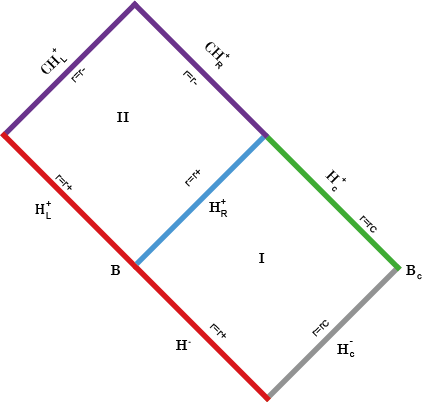}
\caption{Penrose diagram for the RNdS and Kerr-dS black holes (truncated to the physical regions I and II). Each point denotes an $S_2$ ($S_1$) in the case of the RNdS (Kerr-dS black hole).\label{fig:causalstruct}}
\end{figure}

This spacetime is pathological unless $f(r)$ has 3 positive roots. $r_- \leq r_+ \leq r_c $, corresponding to the Cauchy horizon $\mathcal {CH}$, event horizon $\mathcal {H}_R$ and cosmological horizon $\mathcal H_C$. These will bound the different causal regions of the black hole. (see figure \ref{fig:causalstruct}).\\

Region I (the black hole exterior), is bounded by $r_+\,\leq\,r\,\leq\,r_c$. In here $r\,=\,r_+$ corresponds to the white hole horizon ($\mathcal H^-$) at $t=-\infty$ and the right black hole horizon at $t=+\infty$ ($\mathcal H_R^+$). The radial coordinate $r\,=\,r_c$ will correspond to the past ($t=-\infty$) and future ($t=+\infty$) cosmological horizons: ($\mathcal H_C^-$ and $\mathcal H_C^+$).\\

Similarly, region II is bounded by $r_-\,\leq\,r\,\leq\,r_+$. At $r=r_-$ we find the Left ($t=+\infty$) and Right ($t=-\infty$) Cauchy horizons ($\mathcal{CH}_L^+$ and $\mathcal{CH}_R^+$ respectively).  On the other hand, we have in the causal past $r=r_+$, corresponding to the left ($t=-\infty$) and right ($t=+\infty$) future event horizons ($\mathcal H_L^+$ and $\mathcal H_R^+$).\\

In the calculations that follow, we will work in units such that $\Lambda \,= \,3$. Furthermore, it is instructive to use the roots of f to characterize the black hole (instead of M and Q). To do so, we rewrite \eqref{eq:f0}:\\

\begin{equation}
	f(r)\,=\,-\,\frac{\l(r\,-\,r_1\r)\,\l(r\,-\,r_2\r)\,\l(r\,-\,r_3\r)\,\l(r\,-\,r_4\r)}{r^2} \label{eq:f1}\quad.
\end{equation}
where $r_i$ may specify any of the roots of $f$.\\

To make the structure in \eqref{eq:f1} explicit, we take $r^2 f(r)$ in \eqref{eq:f0} and explicitly evaluate the polynomial quotient with respect to $(r\,-\,r_1)$. Then, we divide the result by $(r\,-\,r_2)$. Equating the remainders of this operations to zero, we obtain:
\begin{align}
	&M=\,\frac{1}{2}\,\left(r_1\,+\,r_2\,\r)\,\l(1\,-\,r_1^2\,-\,r_2^2\r) \label{eq:M}\\[.5em]
	&Q^2 =\,r_1\,r_2\,\l(1\,-\,r_1\,r_2\,-\,r_1^2\,-\,r_2^2\r)\label{eq:Qsqr}\\[.8em]
&\!\begin{multlined}
	\frac{r^2f(r)}{\l(r\,-\,r_1\r)\,\l(r\,-r_2\r)}=\label{eq:f2}\\
	r^2\,+\l(r_1\,+\,r_2\r)\,r\,+\,r_1^2\,+\,r_2^2\,+\,r_1\,r_2\,\,-\,1\quad.
\end{multlined}
\end{align}

Finally, equating \eqref{eq:f2} to 0, we find that

\begin{equation}
	r_{3/4} = -\frac{1}{2} \left(r_1\,+\,r_2\pm\,\sqrt{4\,-\,2\, r_1\, r_2\,-\,3\,\l(r_1^2\,+\,r_2^2\r)}\right)\label{eq:r3}\\
\end{equation}

\begin{figure}
\centering
\includegraphics[width=\linewidth]{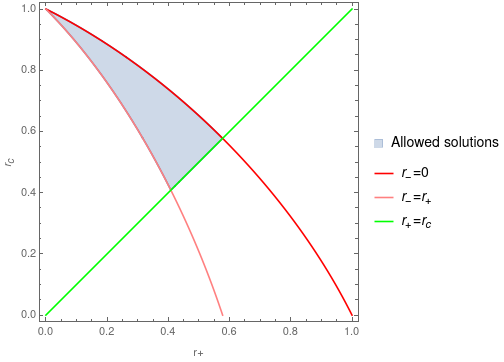}
\caption{Parameter space of the RNdS black hole.\label{fig:RNdSparspace}}
\end{figure}

Lets now define, $r_1\,=\,r_+$, $r_2\,=\,r_c$, $r_3\,=\,r_-$ and $r_4\,=\,r_{\tilde c}$.  Seeking physical solutions, we must impose that the roots $r_i$ are real valued, and obey $0 \leq r_- \leq r_+ \leq r_c $. As we can see in figure \ref{fig:RNdSparspace}, this bounds the allowed RNdS black holes by three limiting cases: $r_-\,\rightarrow \,0$ (Schwarzschild de Sitter limit), $r_-\,\rightarrow \,r_+$ (extremal limit) and $r_+\,\rightarrow \,r_c$ (Nariai limit).\\

Although $r_+,\,r_-$ and $r_c$ are quite useful to discuss properties of the black hole, they are not the best when scanning the BH parameter space (see \ref{fig:RNdSparspace}). Instead, it is more intuitive to consider $Q^2/Q^2_{\text{max}}$ and $y_+ = r_+/r_c$. Here, $Q^2_{\text{max}}$ is the maximal electric charge for a given value of $y_+$. It corresponds to the charge of an extremal RNdS black hole. To obtain this, we take $r_1 \,=\, r_3 $ in equation \eqref{eq:r3}, and solve with respect to $r_1 / r_2$. We get:

\begin{equation}
	Q^2_{\text{max}} = \frac{y_+^2\,\left(2 y_+\,+\,1\right)}{\left(3 y_+^2\,+\,2 y_+\,+\,1\right)^2}\label{eq:Qsqrmax}
\end{equation}

These parameters have the advantage of being defined in the interval $(0,\,1)$, transforming the parameter space into a square. Given equations \eqref{eq:M} and \eqref{eq:Qsqr}, we may invert the definitions of $y_+$ and $Q^2/Q_{\text{max}}^2$ to obtain $r_{+/-/c}$ as a function of $y_+$ and $Q^2/Q_{\text{max}}^2$. \\

In the calculations below, we will also make use of the tortoise coordinate, defined through:
\begin{equation}
	\dd r_* = \frac{\dd r}{f(r)} \label{eq:TortRNdS}.
\end{equation}
$r_*$ goes from $+\infty$ at $r\,=\,r_c$, to $-\infty$ at $r\,=\,r_+$ and back to $+\infty$ at $r\,=\,r_-$. It is then important to specify the  $r$ region, when working with this coordinate. Finally, using ingoing Eddington-Finkelstein coordinates, (see e.g. \cite{WaldBook}) we obtain the surface gravity $\kappa$ of the Horizons:

\begin{equation}
	\kappa_i = \l|\frac{f'(r_i)}{2}\r| \label{eq:SurfGravRNdS}
\end{equation}

\subsection{The Kerr-dS black hole solution\label{subsec:Kerr-dS}}

The Kerr-dS black hole has very similar properties to the RNdS BH. This solution describes a rotating black hole in a de Sitter background, see e.g. \cite{WaldBook}. In Boyer-Lindquist  coordinates, the line element reads:

\begin{multline}
	\dd s^2 = -\frac{\Delta_r}{\Sigma^2\rho^2}\l(\dd t\,-\,a \sin^2\theta\,\dd \phi\r)^2\, +\,\frac{\rho^2}{\Delta_r}\dd r^2\\
	+\frac{\Delta_\theta \sin^2\theta}{\Sigma^2\rho^2}\l(a \dd t-(r^2+a^2)\dd \phi\r)^2+\,\frac{\rho^2}{\Delta_\theta}\dd \theta^2\quad, \label{eq:lineKerrdS}
\end{multline}
where,
\begin{equation}
\begin{split}
	\Delta_r\,&=\l(r^2+a^2\r)\l(1-\frac{\Lambda}{3}r^2\r)\,-\,2\,M\,r,\\
	\Delta_\theta\,&=\,1 + \alpha \cos^2\theta, \\
	\Sigma &= 1 +\alpha,\\
	\rho^2&=r^2 +a^2\cos^2 \theta,\\
	\alpha &= \frac{\Lambda}{3} a^2\quad.
\end{split}
\end{equation}

The black hole is completely specified by the parameters $(M,\,a,\,\Lambda)$, denoting the mass, spin and cosmological constant respectively. As before, the solution is pathological, unless $\Delta_r(r)$ possesses 3 positive roots, $r_- \leq r_+ \leq r_c $, corresponding to the Cauchy, event and cosmological horizon. This condition imposes bounds on the allowed values of the BH parameters, as in the RNdS case. As before, these roots indicate the locations of $\mathcal{CH}$, $\mathcal{H}^+$ and $\mathcal{H}_c$.\\

The causal structure of a Kerr-dS black hole is very similar to the one for a RNdS spacetime. Region I is the exterior of the black hole, bounded by $\mathcal H^+$ and $\mathcal H_C$, while region II is the BH interior, bounded by $\mathcal H^+$ and $\mathcal{CH}$ (see figure \ref{fig:causalstruct}).\\

As before, we may define the tortoise coordinate, by integrating:
\begin{equation}
	\dd r_* = \frac{\Sigma \l(r^2+a^2\r)}{\Delta_r(r)}\dd r \label{eq:TortKerrdS}\quad.
\end{equation}
Finally, we may obtain the surface gravity of the different horizons:
\begin{equation}
	\kappa_i = \frac{1}{2\,a}\frac{\Delta_r'(r_i)}{\Sigma}\,\Omega\l(r_i\r)\quad,
\end{equation}
where:
\begin{equation}
	\Omega(r) = \frac{a}{a^2+r^2} \label{eq:Omega}\quad.
\end{equation}

\section{Quasinormal modes and Strong Cosmic Censorship}
\subsection{Quasinormal modes definition\label{subsec:introdQNM}}
Quasinormal modes have been studied extensively in the literature, see \cite{CardosoQNMs, KokkotasQNMs} for a comprehensive review. In asymptotically de-Sitter spacetimes, QNMs are responsible for governing the late time behaviour of linear fields. We define QNMs as solutions of a given wave equation with specific boundary conditions. Specifically, we require that QNMs are purely outgoing at $\mathcal H_C$, and ingoing into $\mathcal H_R$. \\

As a toy example, let's study QNMs arising from the Klein Gordon (KG) equation in these spacetimes. A massless KG field will obey:
\begin{equation}
	\Box\,\Phi = 0 \label{eq:KG}\quad.
\end{equation}

Quasinormal mode solutions of this equation, can be obtained by performing the \textit{Ansatz}:
\begin{equation}
	\Phi_{n l m}(t,r,\theta,\phi) = e^{-i \omega t} e^{-i m \phi} R_{n l m}(r)\,\Theta_{n l m}(\theta)\quad, \label{eq:KGAnsatz}
\end{equation}
where $\omega$ is the Quasinormal frequency, $n,l\,\text{and}\,m$ are integers that label each mode and $R_{nlm}(r)$ and $\Theta_{nlm}(\theta)$ are the radial and angular part of the solution. In the RNdS black hole, spherical symmetry implies $e^{i m \phi}\Theta_{nlm}(\theta)$ is given by the usual spherical harmonic functions ($Y_{lm}$). In the Kerr-dS background there is no closed form solution known for $\Theta_{nlm}(\theta)$. Plugging equation \eqref{eq:KGAnsatz} into \eqref{eq:KG}, we separate the equation into two ODEs describing the radial and angular part respectively. For convenience, we will henceforth drop the $nlm$ indexes. The radial equation reads:

\begin{equation}
	\l(\dv[2]{}{r_*}\, +\l(\omega\,-\,m\,\Omega_{BH}(r)\r)^2- V_{BH}(r)\r)R(r)\,=\,0 \label{eq:RadialEqGeneral},
\end{equation}
where $V_{BH}$ is the scattering potential, $\Omega_{BH}(r)\,=\,0$ in the RNdS case and $\Omega_{BH}(r)$ is given by equation \eqref{eq:Omega} in the Kerr-dS case. For both BHs, we have $V_{BH}(r_c)\,=\,V_{BH}(r_+)\,=\,V_{BH}(r_-)\,=\,0$.\\

Using a Frobenius analysis, we may define solutions of equation \eqref{eq:RadialEqGeneral} according to their asymptotic behaviour near each horizon. We will denote these solutions according to their ingoing/outgoing character (in/out) at the horizon. For $r\,<\,r_c$, taking the limit $r\rightarrow r_c$ we have:

\begin{equation}
\begin{array}{c c}
	&R_{\text{out},\,c}(r_*) \sim e^{i\,\l(\omega\,-\,m \Omega_{BH} (r_c)\r)\,r_*},\\
	&R_{\text{in},\,c}(r_*) \sim e^{-i\,\l(\omega\,-\,m \Omega_{BH} (r_c)\r)\,r_*},
\end{array}
\qquad \text{as}\qquad r_* \rightarrow +\infty, \label{eq:Rc}
\end{equation}

Similarly, for $r\,>\,r_+$, we define ingoing and outgoing modes such that:

\begin{equation}
\begin{array}{c c}
	&R_{\text{out},\,+}(r_*) \sim e^{i\,\l(\omega\,-\,m \Omega_{BH} (r_+)\r)\,r_*},\\
	&R_{\text{in},\,+}(r_*) \sim e^{-i\,\l(\omega\,-\,m \Omega_{BH} (r_+)\r)\,r_*} 
\end{array}
\qquad \text{as}\qquad r_* \rightarrow -\infty,  \label{eq:Rp}
\end{equation}

The same approach is valid for $r\,<\,r_+$ in the $r\rightarrow r_+$ limit, by taking $r_*$ as a function of $r\in (r_-,\,r_+)$. The asymptotic behaviour of such solutions is identical to the one in \eqref{eq:Rp}. Finally, for $r > r_-$, in the limit $r\rightarrow r_-$, we have:

\begin{equation}
\begin{array}{c c}
	&R_{\text{out},\,-} \sim e^{i\l(\omega\,-\,m \Omega_{BH} (r_-)\r)\,r_*}\\
	&R_{\text{in},\,-}\,\sim\,e^{-i\l(\omega\,-\,m \Omega_{BH} (r_-)\r)\,r_*} 
\end{array}
\qquad \text{as}\qquad r_* \rightarrow +\infty\quad. \label{eq:Rm}
\end{equation}

We can now define quasinormal modes as solutions of \eqref{eq:RadialEqGeneral} that are proportional to $R_{\text{in},\,+}$ and $R_{\text{out},\,c}$:
\begin{equation}
	R(r) \sim R_{\text{in},\,+} \sim  R_{\text{out},\,c}\quad, \label{eq:QNM}
\end{equation}
The condition of proportionality between $R_{\text{in},\,+}$ and $R_{\text{out},\,c}$ is very stringent and quantizes the spectrum of frequencies that lead to QNMs. We call these values of $\omega$ quasinormal frequencies. \\

This condition can be rewritten using the Wronskian W. Given two solutions $f,\,g$ of a linear ODE, $W\l[f,g\r]$ is defined as 
\begin{equation}
	W\l[f,g\r] = f'\,g\,-\,f\,g' \quad.\label{eq:wronskian}
\end{equation}
If f is proportional to g, we have $W[f,g]\,=\,0$. We can thus define pairs (QNM, quasinormal frequencies) as solutions of 
\begin{equation}
	W\l[R_{\text{in},\,+},\,R_{\text{out},\,c}\r]\,=\,0\quad.\label{eq:ExtQNM}
\end{equation}
where $'$ denotes differentiation wrt. $r^*$.

\subsection{Relation with Strong Cosmic Censorship\label{sec:relationSCC}}

The late time behaviour of linear fields in a RNdS or Kerr-dS background can be obtained by taking a linear combination of QNMs. Hence, we may expect these modes to control the regularity of fields at the Cauchy horizon. In fact, SCC violations are fully dependent on the behaviour of some families of QNMs. A proper justification of this is a bit lengthy, and can be found in refs. \cite{HarveyRoughSmooth, HarveyBTZ, HarveyRNdSchargedField, HarveyBTZ}. Nevertheless, we will outline the general idea below.\\

Given a RNdS or Kerr-dS black hole, there will be a violation of SCC, if we may extend \textit{generic} small metric perturbations across the Cauchy Horizon, as a solution of the equations of motion. QNMs describe the late time behaviour of perturbations, so we expect the slowest decaying QNM (lowest negative imaginary part) to control the regularity of \textit{generic} perturbations at $\mathcal {CH}_R$.\\

Black holes created by gravitational collapse do not have a left Cauchy horizon, hence, we are interested in characterizing the behaviour of QNMs at $\mathcal{CH}_R$. For simplicity, we will restrict to the $\Omega_{BH}\,=\,0$ case. The same analysis is valid in the general case by performing an adequate coordinate transformation on $\phi$ (see \cite{HarveySccKerrdS}).\\

We start by defining ingoing and outgoing EF coordinates, $u\,=\,t\,-\,r_*$ and $v\,=\,t\,+\,r_*$. Then, we define Kruskal coordinates in the BH interior:
\begin{equation}
\begin{split}
	U_{-} &=-e^{\kappa_{-}\,u}\quad,\\
	V_{-} &= -e^{\kappa_{-}\,v}\quad.
\end{split}
\end{equation}

Reinstating time-dependence in the radial equation, we obtain:
\begin{equation}
\begin{array}{c c}
	e^{- i\omega\,t}\,R_{\text{in},\,-} \sim V_- ^{\frac{i \omega}{\kappa_-}}\\
	e^{- i\omega\,t}\,R_{\text{out},\,-} \sim U_- ^{-\frac{i \omega}{\kappa_-}}
\end{array}
	\qquad\text{as}\qquad
\begin{array}{c c}
	r_*&\rightarrow +\infty\\
	t&\rightarrow +\infty\quad .
\end{array}
\label{eq:asmympKruskalHp}
\end{equation}
The right Cauchy horizon is obtained by taking the $V \rightarrow 0$ limit, with $U_- > 0$. The outgoing solution is smooth at the horizon, whereas the regularity of the ingoing part is dictated by the value of 
\begin{equation}
	\beta\,=\,-\text{Im}\l(\frac{\omega}{\kappa_-}\r)\quad. \label{eq:beta}
\end{equation}

In fact, if $\beta \,>\,\frac{1}{2}$, $R_{\text{in},-}$ can be continued across $\mathcal{CH}_R$ with finite energy. Similarly, for some positive integer k, if $\beta \geq k$ then $e^{-i \omega t}R_{\text{in},-}(r)$ has $C^k$ regularity at the Cauchy Horizon.\\

At $\mathcal H^+$, QNMs are proportional to $R_{in,+}$. Extending this to the interior of the Black Hole, we may decompose each mode as
\begin{equation}
	R_{\text{in},+}(r) = \mathcal A (\omega)\,R_{\text{in},-}(r)\,+ \mathcal B(\omega)\,R_{\text{out},-}(r). \label{eq:RinpExpansion}
\end{equation}
Given the smoothness of $e^{-i \omega t}R_{\text{out},-}$, the regularity of a QNM at $\mathcal{CH}_R$ should be dictated by the value of $\beta$. We may then argue that the irregular behaviour of our generic perturbation is controlled by the frequency of the slowest decaying QNM. Nevertheless, there is a subtlety we must be wary of. If the slowest decaying QNM happens to be purely ingoing at the Cauchy Horizon, i.e. $\mathcal A (\omega)\,=\,0$, then the mode will be smooth there. The regularity of waves will be dictated by the slowest decaying mode, with \textit{non-vanishing} $\mathcal A (\omega)$. \\

The condition $\mathcal A (\omega)\,=\,0$ can be rewritten as 

\begin{equation}
	W[R_{\text{in},+},\,R_{\text{out},-}]\,=\,0\quad.\label{eq:IntQNM}
\end{equation}

This condition is akin to equation \eqref{eq:ExtQNM}. In fact, it quantizes the spectrum of solutions to \eqref{eq:KG}. We define the modes in this spectrum as interior QNMs \cite{HarveyBTZ}. These solutions will be purely ingoing at  $\mathcal H_+$ and outgoing at $\mathcal{CH}_R$. We conclude that if the lowest lying exterior QNM has the same frequency as onw of the interior modes, the regularity of linear fields at $\mathcal{CH}_R$ will be increased.\\

In \cite{HarveyRoughSmooth, HarveyRNdSchargedField, HarveySccKerrdS}, studies of SCC in the RNdS and Kerr-dS BHs were performed assuming that $\mathcal A (\omega)\,\neq\,0$ generically. On the other hand, in \cite{HarveyBTZ}, Dias et al. show that $\mathcal A (\omega)\,=\,0$, for a family of exterior modes. This coincidence leads to the subsequent SCC violation found in the paper. It would be very interesting if this coincidence was present in the Kerr-dS black hole. The increased regularity of modes at the Cauchy horizon would lead to a SCC violation in the vacuum Einstein Equations, with $\Lambda > 0$. Nevertheless, the authors of \cite{HarveyBTZ} argue that this coincidence is most likely a special case of the 3-dimensional BTZ BH. In fact, the wave equation is hypergeometric in this background, containing only 3 singular points. Contrarily, for the 4 dimensional RNdS / Kerr-dS BHs, the equations of motion are of the Huen type, with 4 or more singular points, with more degrees of freedom. Hence, it is less likely we find a coincidence between interior and exterior QNM frequencies. In this paper we will explicitly compare the frequency spectrum of interior and exterior QNMs, checking the validity of this argument.

\section{Strong Cosmic Censorship in the RNdS black hole \label{sec:SCCRNdS}}
\subsection{Master Equation}

\begin{figure*}
\centering
  \begin{subfigure}[b]{0.4\textwidth}
    \includegraphics[width=\textwidth]{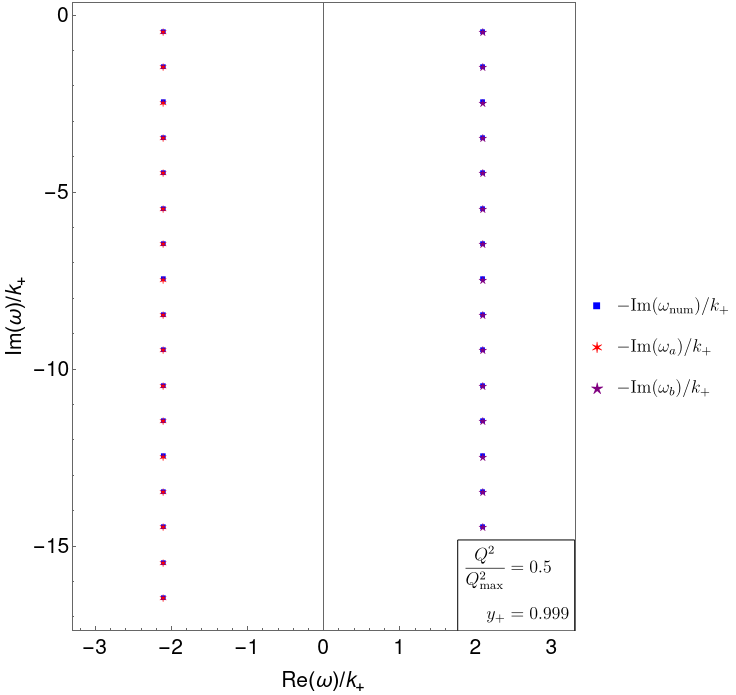}
    \caption{Nariai limite spectrum}
    \label{fig:NariaiModes}
  \end{subfigure}
  \begin{subfigure}[b]{0.4\textwidth}
    \includegraphics[width=\textwidth]{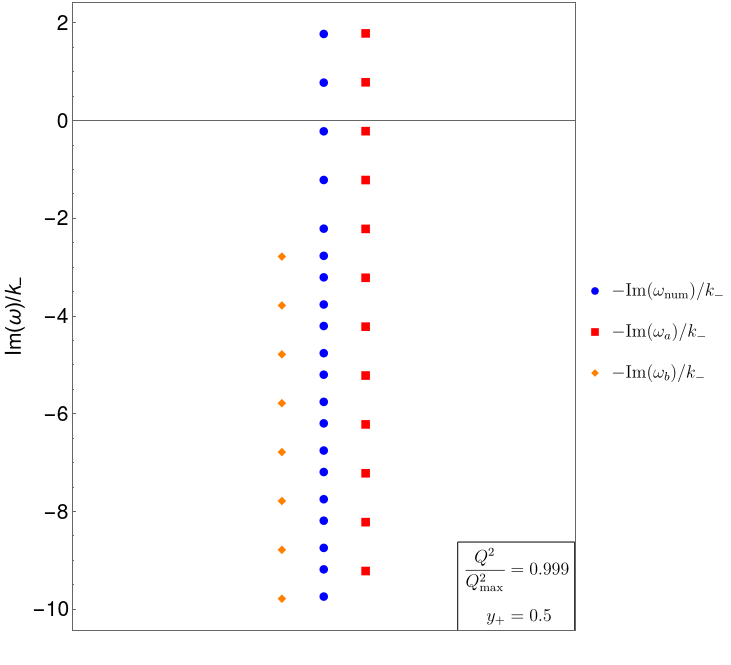}
    \caption{Interior near extremal spectrum}
    \label{fig:InteriorModeSpectrum}
  \end{subfigure}
  \caption{Comparison between the analytical approximations obtained for the Nariai limit (left panel) and the near extremal interior spectrum (right panel) with results obtained numerically. In both cases we studied quasinormal modes of $\Phi_-$ with $l=2$. On the left panel we have $Q^2/Q^2_{\text{max}}\,=\,0.5$, $y_+\,=\,0.999$ and in the right panel we have $y_+\,=\,0.5$, $Q^2/Q^2_{\text{max}}\,=\,0.999$. Notice that in the right panel, the frequencies are purely imaginary, the horizontal displacement is artificial for readability purposes.}
\end{figure*}

For most cases, the equations of motion of linear fields in a RNdS black hole, can be written in the form of the following master equation:
\begin{equation}
	\l(\dv[2]{}{r_*}\,+\,\omega^2- V(r)\r)\,R(r)\, =\, 0 \label{eq:master},
\end{equation}
where $V(r)$ is smooth for $r>0$. In most relevant cases, we can further decompose 
\begin{equation}
	V(r) = f(r)\,A(r), \label{eq:V}	
\end{equation}
where $V(r)$ is smooth for $r>0$. The expression for $V(r)$ in \eqref{eq:master} depends on the specifics of the field we are considering. In this work we will focus in  gravito-electromagnetic perturbations, as studied in \cite{HarveyRoughSmooth}. These modes may be constructed from the combination of two scalar modes ($\Phi^{\pm}_\text s(r)$) and two vector modes ($\Phi^{\pm}_\text v(r)$), obeying equation \eqref{eq:master}, with $V_{\text{v/s},\,\pm}(r)$ given by,

\begin{align}
	V_{\text s,\pm} &= \beta_\pm f(r) \dv{\check F_\pm}{r}\,+\,\beta^2_\pm \check F_\pm(r)^2\,+\,\tilde\kappa \check F_\pm(r)\label{eq:Vs}\quad,\\
	V_{\text v,\pm} &= -\beta_\pm f(r) \dv{\check F_\pm}{r}\,+\,\beta^2_\pm \check F_\pm(r)^2\,+\,\tilde\kappa \check F_\pm(r)\label{eq:Vv}\quad,
\end{align}
where,
\begin{align}
	\beta_\pm &=3 M\,\mp\,\sqrt{9M^2+ 4\,Q^2(l-1)(l+2)}\label{eq:beta},\\[0.5em]
	\tilde \kappa &= (l-1)(l+2)[(l-1)(l+2)+2],\label{eq:kappa}\\
	\check F_\pm &= \frac{f(r)}{r[(l-1)(l+2)r + \beta_\pm]}\label{eq:checkF}.
\end{align}

Here, $l$ is the integer azimuthal quantum number, coming from the expansion in spherical harmonics $Y_{lm}(\theta,\phi)$. For $\Phi^+_ {\text{v/s}}$, $l \geq 1$ whereas for $\Phi^-_ {\text{v/s}}$, $l \geq 2$.\\

In \cite{HarveyRoughSmooth} it was shown that scalar and vector perturbations are isospectral, so we will henceforth drop the $\text{v/s}$ subscript and concern ourselves only with scalar perturbations. It was shown that the lowest lying QNM is either a solution for $\Phi_+$ with $l=1$ or $\Phi_-$, with $l=2$. Below, we will compare the interior QNM spectrum with the frequencies of these two modes.\\

Defining $r_*$ in the appropriate domain, the boundary conditions that define interior and exterior QNMs can be written as:
\begin{equation}
\begin{split}
	&R(r)\,\sim\,e^{i \omega r_*}\quad\text{as}\quad r_* \rightarrow - \infty\,, \\
	&R(r)\,\sim\,e^{-i \omega r_*}\quad\text{as}\quad r_* \rightarrow +\infty \label{eq:BC}\,.
\end{split}
\end{equation}
For interior (exterior) QNM, we denote $r_-\,\l(r_c\r)$ as $r_2$ and $r_+$ as $r_1$. Restoring $r$ dependence in equation \eqref{eq:BC}, we get:
\begin{equation}
\begin{split}
	&R(r)\,\sim\,\l|r\,-\,r_1\r|^{-\frac{i \omega}{2\,k_1}}\quad\text{as}\quad r\,\rightarrow\,r_1 \quad,\\
	&R(r)\,\sim\,\l|r\,-\,r_2\r|^{-\frac{i \omega}{2\,k_2}}\quad\text{as}\quad r\,\rightarrow\,r_2 \label{eq:BCr}\quad.\\
\end{split}
\end{equation}
It is useful to incorporate the boundary conditions in the equations of motion. Thus, we redefine $R(r)$:
\begin{equation}
	R(r)\,=\,\l|r\,-\,r_1\r|^{-\frac{i \omega}{2\,k_1}}\,\l|r\,-\,r_2\r|^{-\frac{i \omega}{2\,k_2}}\, \tilde R(r) \label{eq:tildeR}\quad,
\end{equation}
where $\tilde R(r)$ is smooth in $r_1$ and $r_2$. We now define the dimensionless variable,
\begin{equation}
	y\,=\,\frac{r\,-\,r_1}{r_2\,-\,r_1}\quad,\label{eq:y}
\end{equation}
with $y=0$ at $r_1$ and $y=1$ at $r_2$. Substituting \eqref{eq:y} and \eqref{eq:tildeR} in \eqref{eq:master}, we get
\begin{equation}
	\tilde R''(y)\,+\,\frac{P(y, \omega)}{y\,(y-1)}\,\tilde R'(r)\,+\,\frac{Q(y, \omega)}{y (y-1)}\,\tilde R(y)\,=\,0\label{eq:masterfinal}\quad,
\end{equation}
with $P(y, \omega)$ and $Q(y, \omega)$ analytic functions of $y$ and $\omega$ in $[0,1]\times\mathbb C$. For values of $y > 1$ or $y < 0$ we might find singularities in P and Q. We now define QNMs as solutions of \eqref{eq:masterfinal} that are smooth at $y\,=\,0$ and $y\,=\,1$, with $\tilde R(0), \, \tilde R(1)\neq 0$.

\subsection{Analytical approximation for confluent horizons \label{sec:AnalApprox}}

In \cite{HarveyRoughSmooth}, violations of SCC were found for BHs that were close to extremality. These BHs have $Q^2 \rightarrow Q^2_{max}$ (see equation \eqref{eq:Qsqrmax}), corresponding to $r_- \rightarrow r_+$. The authors found that the slowest decaying QNMs in these BHs are QNMs inherited from the limiting extremal RNdS BHs. These solutions are sharply peaked near $\mathcal H^+$, and vanish quickly for $r > r_+$. We denote them as near extremal QNMs. As argued in \ref{sec:relationSCC} we want to explicitly check if the frequencies of these modes coincide with a subset of the interior QNM spectrum. If frequencies were to coincide, fields would have higher regularity at the Cauchy Horizon.\\

Lets first focus in the more general case of modes propagating between two horizons that are arbitrarily close. This situation describes interior QNMs in the near extremal limit, but also QNMs of near Nariai black holes ($r_+\rightarrow r_c$ limit). We will use the second case as a control on our calculations.\\

Analysing the equation \eqref{eq:SurfGravRNdS}, in the $r_1\rightarrow r_2$ limit, we have $\kappa_i\rightarrow 0$. Given that in \eqref{eq:tildeR} $\omega$ shows up always divided by $k_i$, and the same happens in \eqref{eq:masterfinal} if expanded around $y\,=\,0$, we expect that QNM frequencies will vanish proportionally to $k_i$. Thus, we will work with the dimensionless parameter $\rho\,=\,\frac{\omega}{k_1}$. Due to $r_1$ and $r_2$ being very close, we expect that $A$ (defined in \eqref{eq:V}) will be approximately constant in $r\,\in\,(r_1,\,r_2)$. We expect the same from $F$ defined through
\begin{equation}
	f(r) = (r\,-\,r_1)\,(r\,-\,r_2)\,F(r)\quad. \label{eq:Fdef}
\end{equation}
Hence, a Taylor expansion of \eqref{eq:masterfinal} about $r=r_1$ should yield accurate results. We now define the perturbation parameter $\varepsilon = \frac{r_2-r_1}{r_1}$ , and expand $\tilde R$ and $\rho$:
\begin{equation}
\begin{split}
	\tilde R(r) &= \tilde R_0(r)\,+\,\varepsilon\,\tilde R_1(r)\,+\,...\label{eq:NEExpansion}\\
	\rho &= \rho_0\,+\,\varepsilon\,\rho_1\,+\,...
\end{split}
\end{equation}
Plugging this into equation \eqref{eq:masterfinal},  to zero order in $\varepsilon$, we get:

\begin{multline}
	y(1\,-\,y)\,R_0''(y)\,+\,\l(1\,-\,i\,\rho _0\,-\,2\l(1\,-\,i\,\rho _0\r)y\r)\,R_0'(y)\\
	+\l(\frac{A\l(r_1\r)}{F\l(r_1\r)}\,+\,\rho _0^2\,+\,i\,\rho _0\r)\,R_0(y)\,=\,0\label{eq:0thOrderExp}\quad.
\end{multline}

This is the hyper-geometric equation. In the canonical form, we write this equation as:

\begin{equation}
	y(1-y)w''(y) +\l(c-\l(a+b+1\r)y\r)w'(y)-abw(y)=0\label{eq:hypergeometric}\quad.
\end{equation}

This result is analogous to the one obtained in \cite{SCC2D}, as the near horizon geometry of Black Holes is approximately  $AdS_2$. From \eqref{eq:0thOrderExp} we read off the coefficients a, b and c:

\begin{equation}
\begin{split}
	a&= -i\,\rho_0\,+\,\frac{1-\Delta }{2}\quad,\\
	b&= -i\,\rho_0\,+\,\frac{1+\Delta }{2}\label{eq:hypercoefs}\quad,\\
	c&= 1\,-\,i \,\rho_0\quad.
\end{split}
\end{equation}
where, 
\begin{equation}
	\Delta = \sqrt{1\,+\,\frac{4\,A(r_1)}{F(r_1)}} \label{eq:Delta}\quad.\\
\end{equation}

This result is still valid for $1+4\,A(r_1)/F(r_1)<0$. Here, $\Delta$ will be the root of a negative number, and we must choose a branch to define it. Notice that swapping the branch choice is equivalent to swapping the definition of a and b, so this choice can be arbitrary.\\

The hypergeometric equation is solved by hypergeometric functions. These have been well documented in the literature (see e.g. \cite{BenderBook, Kristensson} ). Following the procedure in \cite{HarveyBTZ}, we prove that requiring smoothness at $y\,=\,0$ and $y\,=\,1$ implies a quantization of a or b. We have that either $a\,=\,-k$ or $b\,=\,-k$ with $k\,=\,0,\,1,\,2,\,...$. Solving this with respect to $\rho_0$ we obtain two families of QNMs, defined as type-a and type-b, with frequencies:

\begin{align}
	\frac{\omega_a}{k_+}\,&=\,-i\l(\frac{1-\Delta}{2}+\,k\r)\,+\,\mathcal O\l(\varepsilon\r)\label{eq:intfreqa}\quad ,\\[0.5em]
	\frac{\omega_b}{k_+}\,&=\,-i\l(\frac{1+\Delta}{2}+\,k\r)\,+\,\mathcal O\l(\varepsilon\r)\quad .\label{eq:intfreqb}
\end{align}

Taking $r_1 = r_+$ and $r_2 = r_c$, we can estimate the spectrum of exterior modes in the Nariai limit ($r_+\rightarrow r_c$). Here $\Delta$ will be purely imaginary as discussed above. $\omega_a$ and $\omega_b$ will have the same imaginary part and symmetric real parts. In figure \ref{fig:NariaiModes}, we compare the spectrum of Nariai modes obtained numerically (see section \ref{sec:NumComp}), with this analytic prediction. These are consistent with the WKB approximation in \cite{HarveyRoughSmooth} and the supplementary material of \cite{CardosoSCC}. \\

Similarly, taking $r_1 = r_+$ and $r_2 = r_-$, we obtain an estimate for the interior frequency spectrum of near extremal black holes. Here, $\Delta$ will be real and positive, with no a-priori upper bound. In fact, there are cases where it is greater than 1, leading to modes with positive imaginary frequencies. This is not cause for concern, as there is no relation between interior QNMs and spacetime  stability (see \cite{HarveyBTZ}). In figure \ref{fig:InteriorModeSpectrum} we compare the analytic prediction with the numerical results obtained in $\ref{sec:NumComp}$.\\

This approach is entirely analogous to the analytical approach used in \cite{HarveyRoughSmooth} to estimate exterior QNM frequencies. In fact, we may extend y to negative values, to  explore the behaviour of $R_0(y)$ in the BH exterior. Taking $R_0(y)$, to be smooth in $y\,=\,0$ and exponentially decaying at $y \rightarrow -\infty$, we get a mode that is smooth in $r_+$ and exponentially decaying for $r> r_+$. These are precisely the conditions required in \cite{HarveyRoughSmooth} to estimate NE modes. It is important to mention that the decaying condition is purely empirical, motivated by the behaviour of solutions observed in numerical calculations. Analysing the solutions of equation \eqref{eq:hypergeometric} in terms of hypergeometric functions, we prove that exponential decay at $y\rightarrow-\infty$ implies smoothness at $y=1$. This means that extending the NE QNMs of \cite{HarveyRoughSmooth} to the interior of the BH, we get interior QNMs. Hence, up to 0th order in $\varepsilon$ the spectrum of exterior NE QNM frequencies is a subset of the spectrum of interior QNM frequencies. In fact, NE QNMs coincide with the type-b family of modes defined in \eqref{eq:intfreqb}. \\

This result might seem a bit worrying, as it hints against the conclusion of our paper. In fact, we must compute higher order corrections to distinguish the interior and exterior spectrum. Doing this analytically is hard, and it is not clear how to proceed. Hence, we will resort to numerics, to pinpoint whether there is a difference between interior and exterior QNM frequencies.

\subsection{Numerical computation \label{sec:NumComp}}
\begin{figure*}[ht!]
\centering
  \begin{subfigure}[b]{0.48\textwidth}
    \includegraphics[width=\textwidth]{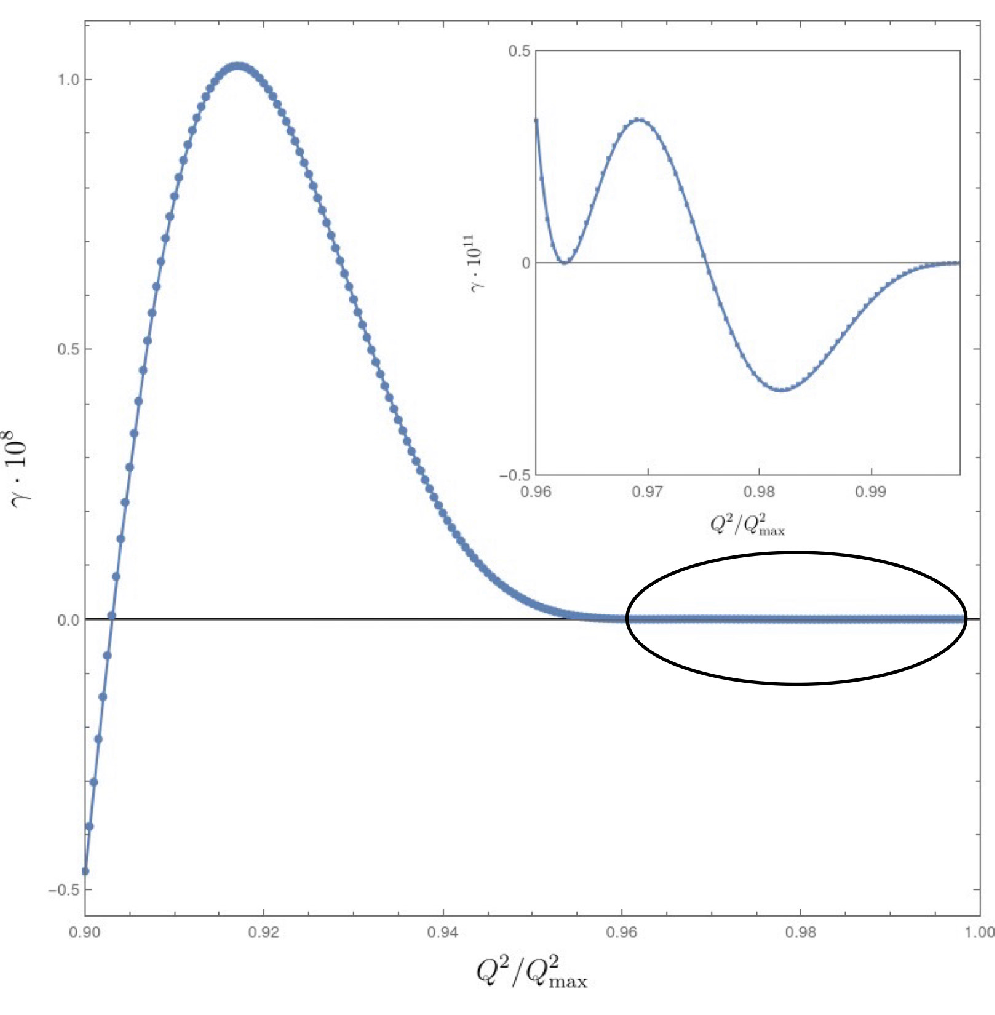}
    \caption{}
    \label{fig:intVsExtLinePlotPhiMinus2}
  \end{subfigure}
  \begin{subfigure}[b]{0.48\textwidth}
  \vskip .cm
    \includegraphics[width=\textwidth]{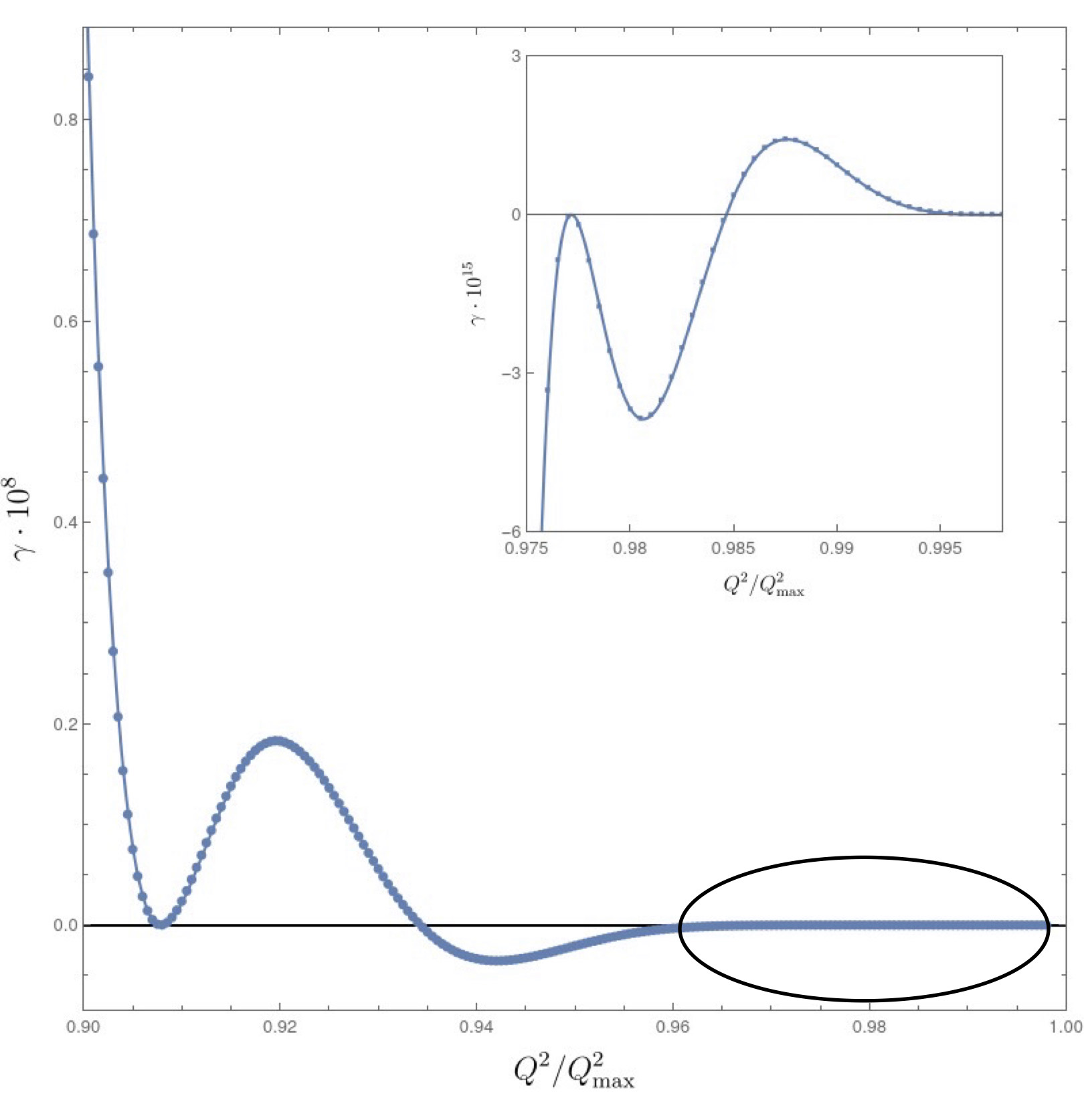}
    \caption{}
    \label{fig:intVsExtLinePlotPhiPos1}
  \end{subfigure}
  \caption{In these images we plot $\gamma$ as a function of $Q^2/Q^2_{max}$ for $y_+ = 0.47$. On the left panel, we study quasinormal modes corresponding to $\Phi_-$, with l=2, whereas on the right panel we have $\Phi_+$, with $l=1$. In both plots, we zoomed the circled region into a subplot on the top right corner. As we can see, the value of $\gamma$ takes very small values, oscillating around 0. We can see that $\gamma$  tends to 0 in the $Q^2\rightarrow Q^2_{max}$ limit, as predicted in section \ref{sec:AnalApprox}.\label{fig:LinePlots}}
\end{figure*}

\begin{figure*}[ht!]
\centering
  \begin{subfigure}[b]{0.48\textwidth}
    \includegraphics[width=\textwidth]{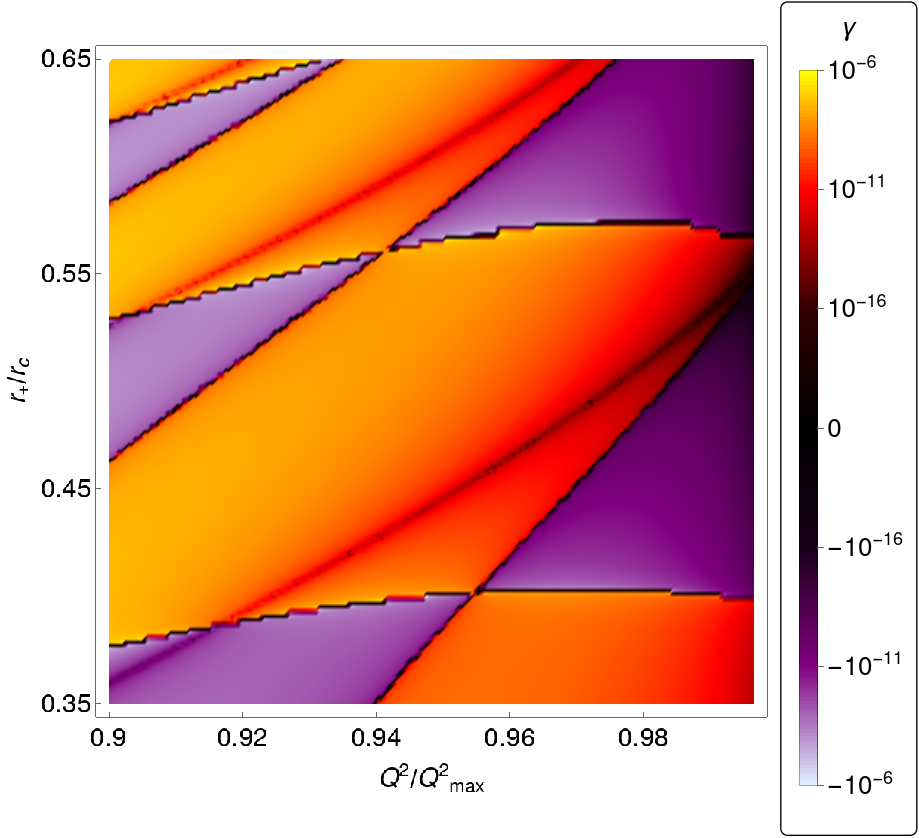}
    \caption{}
    \label{fig:intVsExtDensityPlotPhi2Neg}
  \end{subfigure}
  \begin{subfigure}[b]{0.48\textwidth}
    \includegraphics[width=\textwidth]{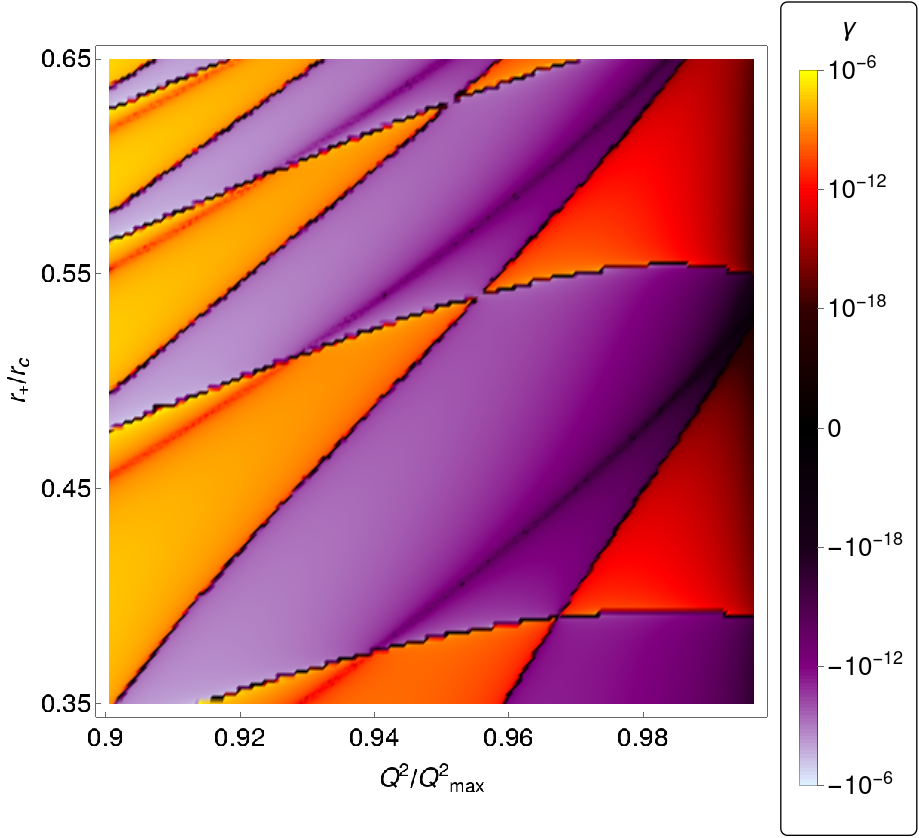}
    \caption{}
    \label{fig:intVsExtDensityPlotPhi1Pos}
  \end{subfigure}
  \caption{As in figure \ref{fig:LinePlots}, we plot the value of $\gamma$ in a non-trivial region of parameter space. On the left panel we study $\Phi_-$, with l=2, whereas on the right panel we have $\Phi_+$, with $l=1$. To produce these plots, we used the Newton-Raphson algorithm to find the frequency of $\omega_{\text{int}}$ and $\omega_{\text{ext}}$ discretely varying the parameter space in a 200x200 grid. As described in section \ref{sec:NumComp}, we repeated the calculation twice for each black hole with different grid sizes, to estimate the numerical precision of our method. We checked that in the whole plot, the numerical precision was always above the difference between interior and exterior frequencies. We remark, that as seen in the one dimensional plot, $\gamma$ goes to 0 as $Q^2\rightarrow Q^2_{max}$.\label{fig:DensityPlots}}
\end{figure*}

To obtain the spectrum of QNMs numerically we approximate the continuous ODE problem by a discrete system of equations, as seen in \cite{SantosNumericalMethods, HarveyHigherDimInstab, SpectralMethodsMatlab, BoydCheb}. More specifically, we sample solutions of equation \eqref{eq:masterfinal} in a discrete Chebyshev grid, mapping the differential operator into a matrix operator. We then solve the resulting matrix equation using standard linear algebra methods.\\

\begin{figure}[ht!]
\centering
    \includegraphics[width=.95\columnwidth]{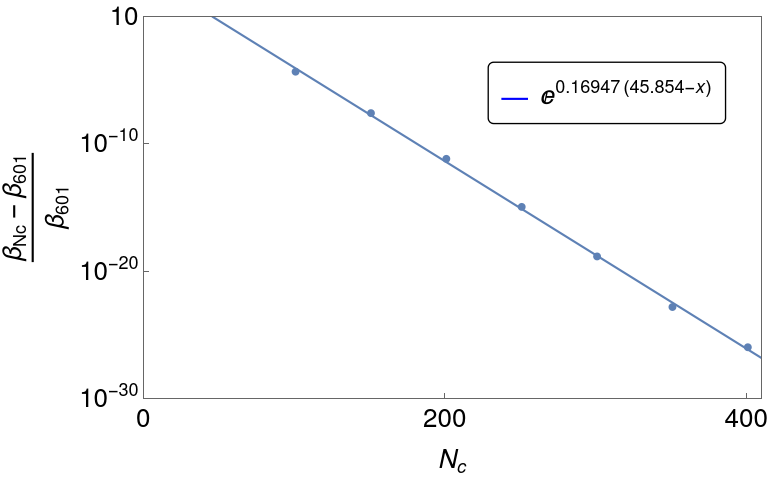}
  \caption{In this figure we show the exponential convergence of the Newton-Raphson method. We plot the normalized difference between $\beta = \omega / \kappa_-$ computed using a given Chebyshev grid size $(N_c)$ and $\beta$ computed using very high resolution $(N_c = 601)$. We see that the relative numerical error decays exponentially with $N_c$ as expected. Here, $\omega$ is the lowest lying NE mode, of $\Phi_-$ with $l=1$, for a BH with $y_+\,=\,0.35$ and $Q^2/Q^2_{max} = 0.99$\label{fig:convergence}.}
\end{figure}

For a given interval $[a,\,b]$ we may define a Chebyshev grid with size $N_c$ as the set of points $x_i$ such that
\begin{equation}
	x_i = \frac{(b-a)}{2} \cos \left(\frac{i\,\pi}{N_c}\right)\,+\,\frac{a+b}{2},\quad \quad i = 0,\,1,\,2,\,...,\,N_c\quad. \label{eq:xc}
\end{equation}
Constraining the domain to the discrete Chebyshev grid, we may rewrite the action of the derivative using matrix multiplication. We denote this matrix $D_c$, (see \cite{SpectralMethodsMatlab} for an explicit definition). Multiplying equation \eqref{eq:masterfinal} by $y(y-1)$ and sampling it in the Chebyshev grid above, we obtain a quadratic matrix eigenvalue problem. The eigenvectors will be discrete approximations of solutions to equation \eqref{eq:masterfinal}, smooth at $y\,=\,0$ and $y\,=\,1$. The eigenvalues will correspond to approximations of the Quasinormal Frequencies.\\

To find the frequencies we use a combination of two methods. We start by obtaining a direct eigenvector decomposition of the system using \textit{Mathematica}'s native \textit{Eigensystem} function. This method yields an approximation of the full frequency spectrum. Bear in mind it is very important to use a software that supports arbitrary precision to obtain good results (we used up to 500 digits of precision in intermediate calculations). This procedure, is not the best to approximate the frequency of individual modes. It is too computational intensive, and does not allow arbitrary high precision of a given frequency. Bearing this in mind, we apply the version of the Newton-Raphson method described in \cite{HarveyHigherDimInstab,SantosNumericalMethods, QuadEigenValue} using the direct method as a seed. Although faster and more precise, this method requires a very accurate initial frequency estimate to converge. In figure \ref{fig:convergence}, we plot the numerical error on the calculation of a given quasinormal frequency as a function of the Chebyshev grid size $(N_c)$. As expected, we see that the usage of a Chebyshev grid actually guarantees that this method has exponential convergence.\\

In the analysis below, we computed the lowest lying NE exterior and type-b interior QNMs (see equation \eqref{eq:intfreqb} for a definition) in a non-trivial region of the parameter space. To achieve this, we used the direct method to obtain these modes for a given RNdS BH, and then used them as seed for a Newton-Raphson method applied to BHs with similar mass and charge. We then used the results as estimates on other similar BHs and henceforth. For each frequency calculated, we repeated this procedure twice, once with a Chebyshev grid size of 500, and then 550. The difference between the obtained frequencies was used to estimate the precision of our calculations. We always made sure this was at least one order of magnitude below the difference between interior and exterior QNM frequencies. This limit would only saturate close to extremality, otherwise, the precision was usually much higher than the difference in frequencies.

\subsection{Comparison between interior and exterior modes for gravitoelectromagnetic perturbations}

As seen in section \ref{sec:AnalApprox}, in the NE limit of RNdS BHs, there is a correspondence between the spectrum of NE exterior modes and type-b interior modes. Using numerics, we verified whether this correspondence extends to the remainder of parameter space. To do so, we explicitly computed the lowest lying frequency of the NE family of exterior modes ($\omega_{\text{ext}}$) and the lowest lying frequency of type-b interior modes ($\omega_{\text{int}}$). To make sure we chose the appropriate families, we compared with the analytic prediction in the near extremal limit (equation \eqref{eq:intfreqb}). We performed this calculation for $\Phi_+\,(l=1)$ and $\Phi_-\,(l=2)$ gravitoeletric QNMs in a large class of RNdS black holes.\\ 

In our plots, we characterize the difference using the quantity 
\begin{equation}
\gamma = \left(\frac{\omega_{\text{in}}\,-\,\omega_{\text{ext}}}{\omega_{\text{in}}\,+\,\omega_{\text{ext}}}\right)\label{eq:gamma}.
\end{equation}
This will always be a real number, as the modes we are concerned with have purely imaginary frequency.\\

In figure \ref{fig:LinePlots} we plot $\gamma$ as a function of $Q^2/Q^2_{\text{max}}$ for a fixed value of $y_+$. Similarly, in figure \ref{fig:DensityPlots}, we depict $\gamma$ in a non-trivial region of parameter space using a density plot. We see that $\gamma$ oscillates around 0, attaining the value in some simple roots and occasional double roots (as seen in the beginning of figure \ref{fig:intVsExtLinePlotPhiMinus2}).\\

\begin{figure}[ht!]
\centering
    \includegraphics[width=.95\columnwidth]{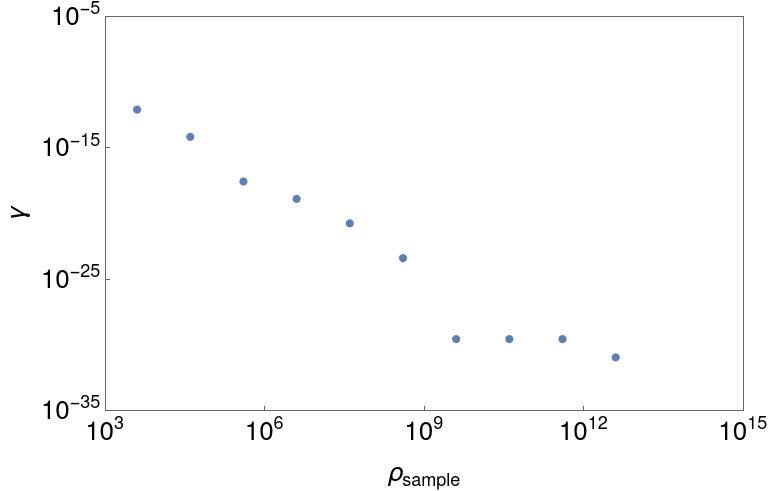}
  \caption{In this figure we study the double roots of $\gamma$ found in figures \ref{fig:LinePlots} and \ref{fig:DensityPlots}. As an example we establish that ($y_+ \approx 0.47$, $Q^2/Q^2_{\text{max}} \approx 0.908$) is indeed a double root of $\gamma$, from $\Phi_+$ QNMs. We fixed $y_+$ and varied $Q^2/Q^2_{\text{max}}$ in a grid of 20 entries, with values in $\overline r_0 \pm \delta_0/2$, ($\overline r_0 = 0.0075$, $\delta_0 = 0.005$). We sampled $\gamma$ there and found the grid point with smallest $\gamma$. Setting this as $\overline r_1$ we created a new grid centred at this point with width $\delta_1 = 10^{-1}\delta_0$. We sampled values of $\gamma$ in this new grid and iterated the process 10 times. In the plot we see the value of the minimum as a function of the sampling density $\rho = 20 / \delta$. The value decreases exponentially with sampling density, hinting that $y_+ \approx 0.47$, $Q^2/Q^2_{\text{max}} \approx 0.908$ is indeed a double root of $\gamma$. \label{fig:DoubleRoot}}
\end{figure}

Analysing figure \ref{fig:DensityPlots}, we see $\gamma$ is always very small, reaching a maximum of $\mathcal O (10^{-6})$. Although the frequencies are very similar as expected, they are not equal. It is interesting to analyse the dark regions in \ref{fig:DensityPlots}. When the colour changes from purple to orange, we have single 0's of $\gamma$, whereas when there is no colour change, we have double zeros, as seen above. In figure \ref{fig:DoubleRoot}, we prove that this features are actually present, by zooming in around one of these roots. These features are unexpected as both the frequencies and their gradient coincide. Close to these lines, there is a wider region where $\gamma$ is close to 0. The roots of $\gamma$ are 1-dimensional regions of parameter space. For these black holes, fields will have increased regularity at the Cauchy horizon. Nevertheless, these regions are not generic in the 2-dimensional parameter space of RNdS BHs, so they are not pointers to SCC violations. Finally, it is worth remarking that $\gamma$ approaches 0 in the extremal limit, validating the discussion in section \ref{sec:AnalApprox}.\\

We can conclude that lowest lying exterior frequencies, \textit{generically}, are not the frequencies of interior modes. Hence the regularity of gravitoelectromagnetic perturbations on a generic RNdS black hole, agree with the ones found in \cite{HarveyRoughSmooth}. Hence, there are no new violations of SCC.\\

\section{Strong cosmic censorship in the Kerr-dS black hole \label{sec:Kerr-dS}}

\subsection{Wave equation and QNM}
As a toy example, we will focus in the behaviour of massless scalar fields in Kerr-dS spacetime. Our argument should be extendible to other fields without much difficulty. In the following, we will show that generic solutions of 
\begin{equation}
	\Box \Phi\,=\,0 \label{eq:wave}
\end{equation}
do not have finite energy at the Cauchy horizon. We will take purely ingoing modes at the event horizon, with the frequency of an exterior mode, and scatter it through the BH interior, studying their regularity at the Cauchy horizon. In a Kerr-dS background, the radial and angular parts of the KG equation are given by the Teukolsky equation. Plugging \eqref{eq:KGAnsatz} into \eqref{eq:KG}, we obtain the following equations:
\begin{align}
	\begin{split}
	\MoveEqLeft[2] {}\dv{}{\theta}\l(\Delta_\theta \sin \theta \dv{\Theta_{\omega l m}}{\theta}\r) \\
	&{}- (m -  a \omega \sin^2\theta)^2 \frac{\Sigma^2}{\Delta_\theta \sin \theta} \Theta_{\omega l m}(\theta) \\
	&{}+ K_{l m}\sin(\theta)\Theta_{\omega l m}(\theta)\,=\,0\quad, \label{eq:TeukolskyAng}
	\end{split}\\[1.5em]
	\begin{split}
	\MoveEqLeft[2] \dv{}{r}\l(\Delta_{r}\dv{R_{\omega l m}(r)}{r}\r)\\
	&+\l[\frac{\Sigma^{2}}{\Delta_{r}}(\omega(r^{2}+a^{2})-am)^2\,-\,K_{l m}\r]R_{\omega l m}(r)=0\quad, \label{eq:TeukolskyRad}
	\end{split}
\end{align}
here $m\in \mathbb{Z},\, |m| \leq l$ and $K_{lm}$ is the constant arising from separation of variables. For ease of notation we will henceforth drop the $\omega l m$ subscript.\\

Equation \eqref{eq:TeukolskyRad} has four regular singular points at the roots $r_i$ of $\Delta_r(r_i)$. For $r>0$ this happens at the Cauchy horizon $(r_-)$, the event horizon $(r_+)$ and the Cosmological horizon $(r_c)$. Solutions to this equation are defined in \eqref{eq:Rc}, \eqref{eq:Rp} and \eqref{eq:Rm}.\\

The analysis of section \ref{sec:relationSCC} tells us that for a QNM with 
\begin{equation}
	\frac{|Im(\omega)|}{\kappa_-} < \frac{1}{2} \label{eq:betaKerrdS}\quad,
\end{equation}
$\Phi_{in,-}$ will have infinite energy at the Cauchy horizon. In \cite{HarveySccKerrdS}, using a WKB analysis, it was proved that taking $l$ large enough we find modes respecting condition \eqref{eq:betaKerrdS}. We will now prove that for these modes the value of $\mathcal A(\omega)$ in equation $\eqref{eq:RinpExpansion}$ is non-zero, validating the conclusions of \cite{HarveySccKerrdS}.

\subsection{Solving the wave equation in the BH interior}

Below, we will study the continuation of an exterior QNM solution, onto the interior of the BH, and show that it has $\mathcal A(\omega) \neq 0$. To do so, we fix $K_{l m}$ and $\omega$ to the values obtained for WKB QNMs defined in \cite{HarveySccKerrdS}. Exterior QNMs are purely ingoing at the event horizon. We use this to set the initial conditions for the wave function in the BH interior. Using a WKB approximation, we solve the radial equation, decomposing it in the $R_{in/out,-}$ basis, using equation \eqref{eq:RinpExpansion}.\\

Consider $r_*$ defined in \eqref{eq:TortKerrdS}, such that $r_*(r_+) =  + \infty$ and $r_*(r_-) =  - \infty$. In terms of $r_*$, the radial equation is 
\begin{equation}
	\dv[2]{S}{r_*}\,+\,U(r,\omega,l,m)\,S(r) \,=\,0 \label{eq:radialS}\quad,
\end{equation}
where $S(r)\,=\,\sqrt{r^2+a^2}\,R(r)$, and
\begin{multline}
	U=\l(\omega-\frac{a m}{r^2+a^2}\r)^2-\frac{K_{l m}\,\Delta_r}{\Sigma^2(r^2+a^2)^2}\\
	+\,(r^2+a^2)^{-4}\Bigg[-\Delta_r^2(r^2+a^2)+3\Delta_r^2r^2\Bigg.\\
	\Bigg.-r\Delta_r\,\dv{\Delta_r}{r}\,(r^2+a^2)\Bigg]\quad .\label{eq:defU}
\end{multline}
The modes studied in \cite{HarveySccKerrdS}, are such that $m=l$. Using a WKB approximation, the authors show that in the large $l$ limit:
\begin{equation}
	\omega = \omega_R + i\,\omega_I =  l\,\Omega_c + \mathcal O (1) \label{eq:WKBomega}\quad,
\end{equation}
here $\Omega_c$ is the rotation frequency of a photon propagating in the inner photon sphere, outside the black hole. This corresponds to $\Omega^+_c$ defined in equation (4.19) of \cite{HarveySccKerrdS}. We expand $K_{ll}$ in powers of $l$:
\begin{equation}
	K_{ll}\,=\,l^2\,\lambda_{-2}\,+\,l\,\lambda_{-1}\,+\,...\label{eq:Kll}
\end{equation}
In appendix \ref{sec:Positivity}, we prove that this expansion is valid and $\lambda_{-2}$ is real and positive.
We can now define:
\begin{equation}
	U_0 = \l[\l(\Omega_c-\frac{a }{r^2+a^2}\r)^2-\frac{\,\lambda_{-2}\,\Delta_r}{\Sigma^2(r^2+a^2)^2}\r]\quad,
\end{equation}
so that
\begin{equation}
	U = l^2 U_0 \,+\, \mathcal O (l) \quad\text{for}\quad l>>1\quad. \label{eq:Uexpansion}
\end{equation}
Notice that $U_0(r_*)$ is a real function of $r_*$, whereas $U(r_*)$ has non vanishing imaginary part. We explicitly verified, that for Kerr-dS BHs:
\begin{equation}
	\Omega(r_-)  > \Omega_c > \Omega(r_+) \label{eq:OmegaHierarchy}\quad,
\end{equation}
where $\Omega(r)$ is defined in \eqref{eq:Omega}.\\

Using equations \eqref{eq:OmegaHierarchy}, \eqref{eq:Kll} and the fact that $\Delta_r(r) < 0$ in $r\in [r_-, r_+]$, we can establish that $U_0(r_*) > 0$ in $r\in [r_-, r_+]$. In the BH exterior $\Delta_r(r) > 0$ and this is no longer the case. \\

We now solve equation \eqref{eq:radialS} using a standard WKB expansion (see \cite{BenderBook}): 
\begin{equation}
	S(r_*) = \exp\l(\frac{1}{l}\sum_{n=0}^\infty \delta^n S_n(r_*)\r).
\end{equation}
Truncating to $n=1$, we obtain that in the large $l$ limit:
\begin{multline}
	S(r_*,\omega,l) =\sqrt{r^2+a^2}\,R_{in,+}\approx\\
	 U_0(r_*)^{-\frac{1}{4}}\Bigg[A(\omega, l)\exp\l(il\int_0^{r_*}U_0(s)^{\frac{1}{2}}\dd s\r)\\
	+ B(\omega, l)\exp\l(-il\int_0^{r_*}U_0(s)^{\frac{1}{2}}\dd x\r)\Bigg] \label{eq:WKBsol}\quad,
\end{multline}
where $A(\omega,l)$ and $B(\omega,l)$ are integration constants.\\

Given that $U_0(r_*) > 0$ everywhere, equation \eqref{eq:WKBsol} is valid everywhere. Physically, this means that to leading order in $l$, there is no scattering of WKB modes propagating in the BH interior. However, to get $\mathcal A (\omega) = 0$ we would need a purely ingoing mode at $\mathcal H^+$ to be fully scattered into an outgoing mode at $\mathcal{CH}_R$. Therefore we can already anticipate that no violations of SCC in the Kerr-dS spacetime will be found. Nevertheless, lets explicitly check this below.

\subsection{Behaviour near the horizons}
For ease of notation, define 
\begin{equation}
	\omega_+ = \omega - l\,\Omega_+\quad,\quad\omega_- = \omega - l\,\Omega_-\quad.
\end{equation}
For $r_*\to -\infty$ (near $\mathcal H_+$), using equations \eqref{eq:defU} and \eqref{eq:WKBomega}, we obtain:
\begin{equation}
	\lim_{r_* \to -\infty} \int_0^{r_*}\l(l \sqrt{U_0} -\omega_+\r)\,=\,\alpha_+\,l\quad,
\end{equation}
where $\alpha_+ = \mathcal O(1)$ as $l\to \infty$. Similarly, for $r_*\to \infty$ (near $\mathcal{CH}_+$), we get
\begin{equation}
	\lim_{r_* \to \infty} \int_0^{r_*}\l(l \sqrt{U_0} -\omega_-\r)\,=\,\alpha_-\,l\quad.
\end{equation}
Dropping the finite term $\sqrt{r^2+a^2}$ in front of $R(r)$, we find that near $\mathcal H_+$,
\begin{multline}
	R(r_*)\sim\l(\frac{\omega_+}{l}\r)^{-\frac{1}{4}}\Bigg[A(\omega, l)e^{-i\l(\omega_+ r_*\,+\,\alpha_+ l\r)} \\
	+ B(\omega, l)e^{i\l(\omega_+ r_* \alpha\,+\,l\r)}\Bigg] \quad \text{as}\quad r_* \rightarrow -\infty\label{eq:WKBsolHplus}\quad.
\end{multline}
Similarly, at $\mathcal{CH}_R^+$, we get:
\begin{multline}
	R(r_*)\sim\l(\frac{\omega_-}{l}\r)^{-\frac{1}{4}}\Bigg[A(\omega, l)e^{-i\l(\omega_- r_*\, +\, \alpha_- l\r)} \\
	+ B(\omega, l)e^{i\l(\omega_- r_* \,+\,\alpha_- l\r)}\Bigg] \quad \text{as}\quad r_* \rightarrow \infty\label{eq:WKBsolCH}\quad.
\end{multline}
Now, we take $R(r_*)$ to be $R_{in,+}$. The ingoing condition (equation \eqref{eq:Rp}) at $\mathcal H_+$ yields $R_{in,+}(r_*)\sim e^{- i \omega_+ r_*}$ as $r_* \rightarrow \infty$. This gives:
\begin{align}
	B(\omega, l) &\sim  \mathcal O (l^{-1})\quad,\\
	A(\omega, l) &\sim  \mathcal O (1)\quad.
\end{align}
Now, focusing in the behaviour near $\mathcal{CH}_R^+$, using equation \eqref{eq:RinpExpansion}, we obtain:
\begin{align}
\mathcal A(\omega) &= \l(\frac{\omega_-}{l}\r)^{-\frac{1}{4}} A(\omega, l) e^{-i\,\alpha_- l}\quad,\\
\mathcal B(\omega) &= \l(\frac{\omega_-}{l}\r)^{-\frac{1}{4}} B(\omega, l) e^{i\,\alpha_- l}\quad,
\end{align}
where 
\begin{equation}
	\frac{\omega_-}{l} = \mathcal O (1)\quad.
\end{equation}
In particular, in large $l$ limit, $\mathcal B$ vanishes, so $\mathcal A(\omega)$ must be non zero, or we would have the trivial solution. Hence, we prove that $\mathcal A(\omega) \neq 0$ for $\omega$ a QMN frequency. The conclusion of  \cite{HarveySccKerrdS} is maintained, preserving SCC for the Kerr-dS BH.

\section{Discussion}
In this paper, we proved that there is no coincidence between the interior and exterior spectrum of QNMs for Kerr-dS and RNdS BHs. Hence, there is no new violation of SCC for these spacetimes. As further work, it would be interesting to relate the discussion in this paper with the quantum instability of Cauchy horizons of Kerr-dS and RNdS black holes, found in \cite{WaldInstabCH}. In fact, this instability is proportional to a factor, that depends on the reflection and transmition coefficients of waves around the BH (see equation 123 of that paper). We suspect that this function could vanish, if interior and and exterior QNMs coincide, akin to $\mathcal A (\omega)$. In fact, in the same paper, the authors study the case of the 3-dimensional BTZ black hole, and they find that due to the hypergeometric nature of the wave equation, this factor vanishes identically in there.\\

\section{Acknowledgements}
The author thanks the contribution of his supervisor Harvey S. Reall, in motivating, discussing and reviewing the contents of this paper, Felicity C. Eperon, for outlining the calculations used in the Kerr-dS case, Jorge E. Santos, Oscar J.C. Dias, Aron Wall, Jonathan Crabbé, Rifath Khan, Bilyana Tomova, Gonçalo Regado and Rita Costa for fruitful physics discussions. The author also thanks Patrícia Jorge for help with producing figure \ref{fig:causalstruct}.  The author is jointly funded by the IOA (University of Cambridge), the University of Cambridge Trust and King's College (University of Cambridge).\\

\appendix
\section{Positivity of $\lambda_{-2}$\label{sec:Positivity}}

In this appendix, we will prove the positivity of $\lambda_{-2}$ defined in \eqref{eq:Kll}. Taking the angular Teukolsky equation \eqref{eq:TeukolskyAng} and making the substitution $x=\cos(\theta)$, we get:

\begin{multline}
	\dv{}{x}\l(1 +\alpha x^2\r)(1-x^2)\dv{}{x} \Theta \\
	-\frac{(m - a \omega (1- x^2))^2\Sigma^2}{(1+ \alpha x^2)(1- x^2)} \Theta  + K_{l m} \Theta \, = \,0\quad.\label{eq:Thetax}
\end{multline}

Multiplying both sides by $\Theta^*$ and integrating in $x \in (-1,1)$, we get:
\begin{equation}
	\int_{-1}^1\Theta^* P_x \Theta\,\dd x + K_{lm} ||\Theta||^2 =0\quad,
\end{equation}
where 
\begin{equation}
	P_x = \dv{}{x}\l(1 +\alpha x^2\r)(1-x^2)\dv{}{x} + \frac{(a \omega (1- x^2)- m)^2\Sigma^2}{(1+ \alpha x^2)(1- x^2)}\quad,
\end{equation}
and 
\begin{equation}
	||\Theta||^2 = \int_{-1}^1 \Theta^* \Theta \,\dd x > 0\quad.
\end{equation}
Integrating by parts the first term of $P_x$ and noting that the boundary terms (evaluation at $x=1$ and $x=-1$) vanish, we get:
\begin{multline}
	K_{lm} = \frac{\int_{-1}^1\frac{(a \omega (1- x^2)- m)^2\Sigma^2}{(1+ \alpha x^2)(1- x^2)} |\Theta|^2\dd x}{||\Theta||^2} \\
	+ \frac{\int_{-1}^1(1 + \alpha x^2)(1-x^2)|\dv{}{x}\Theta|^2\dd x}{||\Theta||^2}\quad. \label{eq:Klm}
\end{multline}
Here, the second term is real and positive, whereas the first term is complex. We have:
\begin{equation}
	K_{lm} = K_{lmR} +  i K_{lmI}\quad.
\end{equation}
However, taking $m = l$ and replacing equation \eqref{eq:WKBomega} in \eqref{eq:Klm}, we obtain:
\begin{equation}
	\frac{K_{lmR}}{K_{lmI}} = \delta\,l + \mathcal O(1)\quad,
\end{equation}
where $\delta$ is some real number. This means that, to leading order in $l$, the coefficients in \eqref{eq:Thetax}, are real. Now, we multiply both sides of \eqref{eq:Thetax} by $(1 +\alpha x^2)(1-x^2)$, and define
\begin{equation}
	\dd x^* = \frac{\dd x}{(1+ \alpha\,x^2)(1-x^2)}\quad,
\end{equation}
with $x^*\in (-\infty, +\infty)$. We rewrite \eqref{eq:Thetax} as:
\begin{equation}
	\dv[2]{}{x_*}\,\Theta(x_*) + Q(x^*)\,\Theta(x^*)\,=\,0\quad,
\end{equation}
with
\begin{multline}
	Q(x_*(x)) = l^2\Bigg((1 + \alpha x^2)(1-x^2) \frac{K_{llR}}{l^2} \\
	- (1-a\,\Omega_c(1-x^2))^2\,\Sigma^2\Bigg) +  \mathcal O (l)\\
	=:l^2\,Q_{-2}(x_*) + \mathcal O(l)\quad.
\end{multline}
We may now study the large $l$ limit of $K_{ll}$, by solving:
\begin{equation}
	\dv[2]{}{x_*}\,\Theta_{-2}(x_*) + l^2\,Q_{-2}(x_*)\,\Theta_{-2}(x_*)\,=\,0\quad.
\end{equation}
Using the fact that $a \Omega_c < 1$ (see \cite{HarveySccKerrdS} equation 4.19) and $\alpha\,<\,1$, we obtain that $Q_2(x_*)$ has 0, 1, or 2 roots. Now, the angular equation must be regular at $\theta\,=0\,$ and $\theta\,=\,\pi$. This is the statement that $\Theta(x_*)$ must decay exponentially at both infinities, i.e. an ingoing wave at $\theta\,=\,0$ must be fully scattered into $\theta\,=\,\pi$. Using a standard WKB approximation (see \cite{BenderBook}) we see that this is the case if
\begin{equation}
K_{llR} \,=\, K_{ll} + \mathcal O(l) \,=\, l^2 \lambda_{-2} + \mathcal O (l).
\end{equation}
Hence, we prove equation \eqref{eq:Kll}. Now, the positivity of $\lambda_{-2}$ follows simply by the substitution of equations \eqref{eq:WKBomega} and \eqref{eq:Kll} into equation \eqref{eq:Klm}. We get:
\begin{multline}
	K_{ll} = l^2 \lambda_ {-2} = l^2\,\Bigg(\frac{\int_{-1}^1\frac{(a \Omega_c (1- x^2)- 1)^2\Sigma^2}{(1+ \alpha x^2)(1- x^2)} |\Theta|^2\dd x}{||\Theta||^2} \\
	 +\,\frac{\int_{-1}^1(1 + \alpha x^2)(1-x^2)|\dv{}{x}\Theta|^2\dd x}{l^2||\Theta||^2}\Bigg) + \mathcal O (l)\quad,
\end{multline}
where both leading order terms are real and positive. Hence, we we deduce that $\lambda_{-2} > 0$.

\bibliography{references}

\end{document}